\documentclass[12pt]{iopart}

\begin{document}

\title{An inhomogeneous fractal cosmological model}

\author{Fulvio Pompilio   
\dag
and Marco Montuori
\ddag
}

\address{\dag\ SISSA (ISAS),Via Beirut 2-4, I--34013 Trieste, Italy}

\address{\ddag\ INFM, Roma1, Physics Department, University "La Sapienza",
P.le Aldo Moro 2,\\ 
I--00185 Roma, Italy}  
\ead{\dag\ pompilio@sissa.it, present address: fulvio@pompilio.roma.it,\\
\ddag\ marco@pil.phys.uniroma1.it}

\begin{abstract}

We present a cosmological model in which the metric allows for an
inhomogeneous Universe with no intrinsic symmetries (Stephani models), 
providing the ideal features to describe a fractal distribution of
matter. Constraints on the metric functions are derived using the
expansion and redshift relations and allowing for scaling number
counts, as expected in a fractal set. The main characteristics of such a 
cosmological model are discussed.

\end{abstract}


\section{Introduction}

One of the most interesting results of observational cosmology over the
last two decades has been the discovery of large scale structures and
voids in matter distribution.\\
To date, however, a clear definition of an homogeneity
scale in matter distribution is still lacking.
Moreover, this controversy concerns 
the morphological features of the galaxy clustering, which
has been found by several authors to have fractal characteristics at 
least up to a certain scale (Sylos Labini, Montuori \& Pietronero 1998; 
Wu et al. 1999). In this case, a suitable cosmological model is lacking, 
given that the Cosmological Principle cannot be applied.\\
Many authors have attempted to develop alternative models, following two
different approaches. The first relies on a perturbative
scheme, through the superposition of a fractal density perturbation to the
FRW background, both in Newtonian gravity (Joyce et al. 2000) and in 
general relativity (e.g. Mittal \& Lohiya 2001). The second concerns 
a more general relativistic study of the metric, used to
match a fractal number counts relation, and 
extensive analytical (e.g. Humphreys et al. 1998)
and numerical (Ribeiro 1992) computations have been performed.\\
In any case, for both approaches,
only Lemaitre-Tolman-Bondi (LTB) inhomogeneous metric has been
considered, i.e. spherically symmetric solutions to Einstein field
equations.\\
Even if this is certainly a very interesting study, it is not 
the most suitable tool to study a fractal distribution.\\
As a matter of fact, in a fractal set of points the number density decreases
on average from any point of the set. In a spherically symmetric
distribution around a point, this is true only if you measure the density 
centering on that point.\\
In this paper, the above restriction is relaxed and an inhomogeneous metric
with no intrinsic symmetries is presented.\\
The starting metric is recovered from the class of Stephani solutions 
(Stephani 1967). Specifically, the aim of this work is to extend the
analytical characterization of a fractal matter distribution to a
Stephani Universe, whose geometrical (e.g Krasinski 1983) and thermodynamical
(e.g. Quevedo \& Sussmann 1995) description has already been widely
discussed and will not be pursued here.\\
In Section 2 the general properties of the Stephani model are recalled
while the consequences on the redshift expression and expansion are presented 
in Section 3. In Section 4 we derive the fractal constraints on the metric 
functions and the main results are discussed in Section 5.\\
Throughout the paper we have assumed natural units ($G=c=1$) and the metric 
signature (+,-,-,-). The Greek indices refer to the whole components,
while the Latin letters are just for the spatial part.

\section{The model}

The metric of the Stephani Universe can be written as follows (Krasinski 1997):
\begin{equation} 
d{s}^{2}={D}^{2}d{t}^{2}-\frac{{R}^{2}(t)}{{V}^{2}}(d{x}^{2}+d{y}^{2}+d{z}^{2})
\end{equation}
where :
\begin{equation} 
V=1+\frac{1}{4}k(t)\{{[x-{x}_{0}(t)]}^{2}+{[y-{y}_{0}(t)]}^{2}
+{[z-{z}_{0}(t)]}^{2}\}
\end{equation}
\begin{equation} 
D=F(t)\left( \frac{\dot{V}}{V}-\frac{\dot{R}}{R}\right) =F\frac{R}{V}
\frac{\partial}{\partial t}\frac{V}{R}
\end{equation}
\begin{equation} 
k(t)=[{C}^{2}(t)-1/{F}^{2}(t)]{R}^{2}(t)
\end{equation}
$C,F,R,{x}_{0},{y}_{0},{z}_{0}$ are arbitrary functions of time and 
dots denote time derivative .\\
Eq.(4) is equivalent to the Friedmann equation, although it must be 
stressed that the analogy is only formal and dictated by the same derivation
(through Einstein field equations), but the function $R(t)$ is not the
equivalent of the scale factor $a(t)$ in FRW models.\\
As can be seen from eq.(1) and more precisely from the expression for
V in eq.(2), the model is inhomogeneous and has no 
intrinsic symmetries, unless ${x}_{0},{y}_{0},{z}_{0}$ are strictly 
constant. In this case ${x}_{0},{y}_{0},{z}_{0}$ 
become the center of spherical symmetry, even if 
the metric is still not homogeneous.\\
Moreover, the distinction between open and closed Universe is more subtle
than in FRW; indeed $k(t)$ is a function of time and so it is not 
fixed as in FRW model. However, it is worth remembering that 
fixing $k(t)$ is not a general requirement of 
Einstein's theory of gravitation.\\
Let us consider a perfect fluid; the stress-energy tensor is:
\begin{equation}
{T}_{\mu \nu}=(\rho+p){u}_{\mu}{u}_{\nu}-p{g}_{\mu \nu}.
\end{equation}
The Stephani metric (Eq.(1)) fulfills 
Einstein field equations if the matter fluid has no 
shear ($\sigma =\frac{1}{2}{\sigma }_{\mu \nu }{\sigma }^{\mu \nu
}=0$), 
no rotation ($\omega =\frac{1}{2}{\omega }_{\mu \nu }{\omega }^{\mu \nu }
=0$) and mass is flowing along t-coordinate lines with:
\begin{equation}
{u}^{\mu}=\frac{V}{F}\frac{R}{R\dot{V}-V\dot{R}}{\delta}^{\mu}_{0}
\end{equation}
\begin{equation}
\dot{{u}^{i}}=\frac{{V}^{2}}{D{R}^{2}}{D}_{,i}~,~\dot{{u}^{0}}=0
\end{equation}
\begin{equation}
\theta\equiv {u}^{\mu}_{;\mu}=-\frac{3}{F}
\end{equation}
where ${\delta}^{\mu}_{\nu}$ is the Kronecker symbol.
$u^{\mu}$ is the $4-velocity$, the notation ${,}_{i}$ means the 
spatial derivative and ${;}_{\mu}$ the covariant derivative.
In the FRW model, $\theta= 3H$ where $H$ is the {\it Hubble constant}.\\
It is important to note that in the model the matter does not flow
along geodesics, as follows from eq.(7); this is a significant
difference with FRW model (which have $\dot{{u}^{i}}=0$) (Krasinski 1983).\\
Turning to thermodynamics properties, Einstein field equations with a perfect
fluid source, the conservation law (${T}^{\mu\nu}_{;\nu}=0$) and
the contracted Bianchi identities 
coupled to the matter conservation equation (${n{u}^{\mu}}_{;\mu}=0$) can be
reconciled in an equation of state, if the thermodynamical quantities
are defined as follows:
\begin{equation}
8\pi \rho(t)=3{C}^{2}(t)
\end{equation}
\begin{equation}
8\pi p(t)=-3{C}^{2}(t)+2{C}^{2}(t){C}_{,t}
\frac{V}{R}{\left( \frac{\dot{V}}{V}-\frac{\dot{R}}{R}\right)}^{-1}
\end{equation}
\begin{equation}
T=\frac{3(p+\rho)}{{n}^{4/3}}
\end{equation}
\begin{equation}
S={S}_{0}+\sigma,\sigma\equiv-\frac{F}{V}
\end{equation}
satisfying the Gibbs-Duhem equation (Quevedo \& Sussman 1995):
\begin{equation}
dS=\frac{1}{T}\left[ d\left( \frac{\rho }{n}\right) +p~d
\left( \frac{1}{n}\right) \right] . 
\end{equation}
As underlined by other authors (Krasinski, Quevedo \& Sussman 1997),
the physical meaning of the equation of state restricts the freedom in
choosing the metric functions, though a defined correspondence among
their properties and the physical quantities cannot be set, in general.
However, since this work is mainly focused on a fractal model characterization,
we will assume that the thermodynamic scheme holds and the metric constraints
will be derived independently.\\ 
Again, a deep difference with respect to FRW models emerges, namely 
the equation of state is not barotropic, as it is evident from Eq.(9)
and Eq.(10).\\
The fact that the equation of state depends on position, as indicated by 
eq.(9)-(12), and the absence of any
intrinsic symmetry make the model particularly suited in the
description of a fractal distribution of matter. 

\section{The redshift equation and the expansion flow}

By definition, the redshift $Z$ is 
determined along a photon path (written
in terms of the affine parameter $\lambda$) as follows (Ellis et al. 1985):
\begin{equation}
(1+Z)\equiv \frac{{({u}^{\mu}{k}_{\mu})}_{em}}{{({u}^{\mu}{k}_{\mu})}_{obs}}=
{\left( \frac{dt}{d\lambda}\right) }_{em}/
{\left( \frac{dt}{d\lambda}\right) }_{obs}.
\end{equation}
On the other hand, the definition of redshift refers to the ratio of the
time shift along observer's world line with respect to the shift along a
general world line, namely:
\begin{equation}
(1+Z)\equiv \frac{dt}{d\tau}.
\end{equation}
A more formal manner to find the aforementioned relation is given by 
considering the photon wave vector ${k}^{\mu }$, which satisfies:
\begin{equation}
{k}^{\mu }\equiv {t}_{,\mu}=\frac{d{x}^{\mu }}{d\lambda}
\end{equation}
\begin{equation}
{k}^{\mu }{k}_{\mu }=0
\end{equation}
therefore, with no loss of generality, normalizing with respect to the
observer's value, it yields:
\begin{equation}
(1+Z)={u}^{0}_{em}=\frac{1}{D}=\frac{V}{F}\frac{R}{R\dot{V}-V\dot{R}}.
\end{equation}
It is worth noting that the above \textit{redshift equation} is
dependent from the position,
which is another important difference with respect to FRW models.\\
In addition, other authors (Ribeiro 1992) have noted that more information 
about photon motion can be gained by evaluating the Lagrangian of the metric 
in eq.(1) and then solving for the Euler-Lagrange equations:
\begin{equation}
\frac{d}{d\lambda }\frac{\partial L}{\partial\dot{{x}^{\mu}}}-
\frac{\partial L}{\partial {x}^{\mu}}=0~,~
\dot{{x}^{\mu}}=\frac{d{x}^{\mu}}{d\lambda}.
\end{equation}
Inserting the past light cone expression relating the derivatives of the
time and spatial parts of the metric (as follows by setting $d{s}^{2}=0$)
into the Euler-Lagrange equation for the time component, a few algebra 
leads to the time motion of photons:
\begin{equation}
\frac{{d}^{2}t}{d{\lambda}^{2}}+\left[\frac{\dot{D}}{D}+\frac{\dot{D}}{{D}^{3}}
-\frac{\dot{R}}{R{D}^{2}}+\frac{\dot{V}}{V{D}^{2}}\right]
{\left( \frac{dt}{d\lambda}\right) }^{2}=0.
\end{equation}
In the motion of a photon the affine parameter $\lambda$ is equivalent to
the proper time $\tau$, so that, from eq.(15) and eq.(18), eq.(20) gives
the \textit{redshift constraint} on the metric:
\begin{equation}
\frac{\dot{D}}{D}+\frac{\dot{V}}{V}-\frac{\dot{R}}{R}=0
\end{equation}
or equivalently:
\begin{equation}
\frac{\dot{D}}{{D}^{2}}+\frac{1}{F}=0.
\end{equation}
Comparing the $V,R,F$ dependence in the redshift equation (eq.18) with
the expansion expression in eq.(8) and including the redshift constraint
(eq.(21)), yields:
\begin{equation}
\frac{\theta}{3}=-\frac{1}{(1+Z)}\frac{\partial}{\partial \tau}(1+Z).
\end{equation} 
The left-hand side of the above equation is the equivalent of the Hubble
parameter in the FRW models ($3H=\theta$). In this case, the corresponding 
relation is: 
\begin{equation}
{H}_{FRW}\equiv \frac{\dot{a}}{a}=-\frac{1}{(1+Z)}\frac{d}{dt}(1+Z)
\end{equation} 
where $a(t)$ is the scale factor.\\
Though a close correspondence between the two expression is apparent, they
are quite different. In the case of an inhomogeneous model, the expansion is 
linked to the matter distribution since $(1+Z)$ (eq.(18)) depends on the
position via the function $V$.\\
In other words, different regions of the Universe can undergo different
expansion histories.\\
This is evident from the expansion law, which appears to
contain an additional term. Indeed, 
the general redshift-distance relation is (Ehlers 1993):
\begin{equation}
Z=\left(\frac{\theta}{3}+{\sigma}_{\mu \nu }{e}^{\mu }{e}^{\nu }\right) 
\delta l-{\dot{u}}_{\mu }{\delta }_{\perp }{x}^{\mu }
\end{equation} 
where ${e}^{\mu }$ is the unit vector joining the observer and the emitting
source, whose distance is $\delta l$, and directed towards the 
source. The position vector ${\delta }_{\perp }{x}^{\mu}$ is defined in the
following way. Let ${h}^{\mu }_{\nu }$ be the tensor projecting the tangent 
vector-space at each point perpendicularly onto the three dimensional 
subspace orthogonal to ${u}^{i}$, namely:
\begin{equation}
{h}^{\mu }_{\nu }={\delta}^{\mu }_{\nu }+{u}^{\mu }{u}_{\nu }.
\end{equation}
Considering a displacement ${\delta x}^{\nu}$, we can define the position 
vector as:
\begin{equation}
{\delta }_{\perp }{x}^{\mu}={h}^{\mu }_{\nu }{\delta x}^{\nu}
\end{equation} 
and then:
\begin{equation}
\delta l={({g}_{\mu \nu }{\delta }_{\perp }{x}^{\mu }{\delta }_{\perp }
{x}^{\nu })}^{1/2}.
\end{equation}
The shear contribution vanishes in the present model, therefore, by using 
the \textit{redshift constraint} and introducing the acceleration in terms
of the metric functions (eq.(7)), eq.(25) can be written as:
\begin{equation}
Z=\frac{\dot{D}}{{D}^{2}}\delta l+\frac{{D}_{,i}}{D}{\delta x}^{i}.
\end{equation}
where $\frac{{D}_{,i}}{D}{\delta x}^{i}$ is the kinematic acceleration
evaluated at the position fixed by $\delta l$.\\
The first term is the analogous to FRW Hubble term (albeit it depends on the
position), since $\frac{\dot{D}}{{D}^{2}}=H$ and $\delta l$
is the comoving line element (see next section, eq.(40)). On the other hand, 
the additional second term is a new dipolar one, that is purely cosmological,
i.e. not due to peculiar velocities.\\  
A few consequences of this will be further analyzed in the 
last section, nevertheless, it is worth noting that the present model 
is suitable to describe a fractal distribution, since all the cosmological 
parameters depend on the position.\\
Using eq.(18) and eq.(29) , the metric
function $D$ must satisfy the following differential equation:
\begin{equation}
\frac{\dot{D}}{{D}^{2}}\delta l+\frac{{D}_{,i}}{D}{\delta x}^{i}
-\frac{1}{D}+1=0
\end{equation}
and its solution can provide the space-time expression for the redshift.\\
Eq.(30) basically describes a nonlinear wave motion with a damping factor. For
instance, if the one dimensional equivalent is considered, eq.(30) has the
general shape of the so-called \textit{Burger's equation}, which is an
approximate description of one dimensional turbulence, with a damping term. 

\section{Fractal constraints}

The density of sources along the $\delta l$ direction after an affine parameter
displacement ($d\lambda $) at some point $P$ is (Ribeiro 1992):
\begin{equation}
{d}^{2}N={\delta {l}_{0}}^{2}d{\Omega}_{0}{[n(-{k}^{\mu}{u}_{\mu})]}_{P}
d\lambda
\end{equation}
where $\delta {l}_{0}$ is the observer area distance 
($\delta {l}_{0}^{2}=\frac{d{S}_{0}} {d{\Omega}_{0}}$) and $n$ is the density 
of radiating sources in a subtended angle $d{\Omega}_{0}$.\\
The density of sources can be found through the proper luminous matter 
density, assuming that the basic structures are galaxies with nearly the same 
mass ${M}_{G}$:
\begin{equation}
n=\frac{{\rho}_{m}}{{M}_{G}}=\frac{3{C}^{2}(t)}{8\pi{M}_{G}}.
\end{equation}
Using eq.(31) and eq.(32), a comoving observer measures a 
number of sources corresponding to:
\begin{equation}
{N}(\delta {l}_{0})=\frac{3{\delta {l}_{0}}^{2}}{2{M}_{G}}
\int_{\Delta t}{{C}^{2}(t)dt}.
\end{equation}
where integration is performed over the time $\Delta t$ spent by the photon
along its path; on the other hand, as a more physical interpretation, it 
directly links the motion of the photon to the intervening matter 
distribution, as the expression in eq.(9) for ${C}^{2}(t)$ shows.\\
This value can be matched to the expected fractal value in the following way.\\
Averaging over a sphere centered at the observer location, a volume and a 
volume density can be defined:
\begin{equation}
V({d}_{l})=\frac{4}{3}\pi {d}_{l}^{3}
\end{equation}
\begin{equation}
{\rho}_{v}({d}_{l})=\frac{{M}_{G}{N}(\delta {l}_{0})}{V({d}_{l})}
\end{equation}
where this time the luminosity distance ${d}_{l}$ corresponding to 
$\delta {l}_{0}$ was introduced, since it is the observational distance 
measure. It can be related to the observer area distance 
by means of the redshift factor $(1+Z)$ through the following:
\begin{equation}
{d}_{l}={\left( \frac{d{S}_{0}}{d{\Omega}_{0}}\right) }^{1/2}{(1+Z)}^{2}=
\delta {l}_{0}{(1+Z)}^{2}.
\end{equation}
A fractal distribution of matter is characterized by a power-law 
scaling\footnote [1] {This scaling law should strictly apply only in an 
Euclidean space-time, since no general relativistic extension to it has been 
derived, so far. However, it has gained observational support, so it is used 
here without any theoretical speculations.}:
\begin{equation}
N({d}_{l})={N}_{0}{\left( \frac{{d}_{l}}{{l}_{h}}\right) }^{3-\gamma }
\end{equation}
\begin{equation}
{\rho}_{v}({d}_{l})=\frac{3{N}_{0}{M}_{G}}{4\pi {l}_{h}^{3}}
{\left( \frac{{d}_{l}}{{l}_{h}}\right) }^{-\gamma }
\end{equation}
where ${l}_{h}$ is a transition scale corresponding to homogeneity
and ${d}_{f}=3-\gamma$ is the fractal dimension.\\
From the above equations and using the measured number counts, it yields:
\begin{equation}
{I}_{\gamma}(\delta {l}_{0},Z)\equiv \int_{\Delta t}
{{C}^{2}(t)dt}
=\frac{2}{3}\frac{{N}_{0}{M}_{G}}
{{l}_{h}^{2}}{\left( \frac{\delta {l}_{0}}{{l}_{h}}\right) }^{1-\gamma }
{(1+Z)}^{2(3-\gamma)}
\end{equation}
which links the unknown metric functions with observable quantities 
${N}_{0}, {M}_{G}, \gamma, {l}_{h}$ and is
the \textit{fractal constraint} on the metric. According to eq.(9), the
metric function $C(t)$ is related to density evolution $\rho (t)$, so that
the physical meaning of the fractal constraint is to relate redshift and
matter distribution in the proper way over each $t=const$ hypersurface
at each position.\\ 
Evaluation of the metric terms at the homogeneity scale 
(${I}_{\gamma}({l}_{h},Z({l}_{h}))$) simplifies the above relations and 
involves the knowledge of the fractal parameters ($\gamma,{l}_{h}$),
the number of galaxies ${N}_{0}$ and the 
typical galactic mass (${M}_{G}$), which are provided by the analysis
of the galaxy catalogues, and the redshift factor, whose expression can be 
inserted from the solution of eq.(30).\\
Otherwise, the value of $\delta {l}_{0}$ must be obtained using the 
proper-distance aperture of the subtended area, which follows from the line
element for the metric in eq.(1). The line element along the $i$-th
axis at a fixed timed ${t}_{0}$ is therefore:
\begin{equation}
\delta {x}^{i}_{0}({t}_{0})=\int_{0}^{\delta {x}^{i}_{0}}
{\frac{R({t}_{0})}{V({t}_{0},{x}^{i}_{0})}d{x}^{i}}
\end{equation}
and can be found as (Krasinski 1983):
\begin{eqnarray}
2R({t}_{0}){[{k}({t}_{0})]}^{-1/2}
\arctan\left( \frac{1}{2}{[{k}({t}_{0})]}^{1/2}
\delta {x}^{i}_{0}\right)&~~k({t}_{0})> 0\\
R({t}_{0})\delta {x}^{i}_{0}&~~k({t}_{0})=0\\
\frac{R({t}_{0})}{{[\mid k({t}_{0})}\mid ]^{1/2}}\ln 
\frac{1+\frac{1}{2}{[\mid k({t}_{0})}\mid ]^{1/2}\delta {x}^{i}_{0}}
{1-\frac{1}{2}{[\mid k({t}_{0})}\mid ]^{1/2}\delta {x}^{i}_{0}}&~~
k({t}_{0})< 0.
\end{eqnarray}
In addition, the value of ${N}_{0}$ can be obtained by setting the matching 
condition to a FRW metric as ${l}_{h}$ is approached (Humphreys et al. 1998), 
i.e. by equating the expression for ${N}(\delta l)$ evaluated at ${l}_{h}$ 
with the FRW analogous expression at the corresponding radial comoving
homogeneity scale ${r}_{h}$ (Coles \& Lucchin 1995):
\begin{equation}
{N}_{FRW}({r}_{h},{k}_{FRW})=4\pi \int_{0}^{{r}_{h}}
{\frac{n[t(r)]a[t(r)]{r}^{2}}{{(1-{k}_{FRW}{r}^{2})}^{3}}dr}
\end{equation}
in terms of the FRW scale factor $a(t)$ and curvature constant 
${k}_{FRW}=-1,0,+1$. The above relation can be expanded and restated using 
the number counts continuity ($n{a}^{3}={n}_{0}{a}_{0}^{3}$), if the
number of particles in the lapse time ${t}_{0}-t(r)$ is kept fixed
(i.e. no particle creation/destruction and evolution), as 
follows (Coles \& Lucchin 1995):
\begin{equation}
{N}_{FRW}({r}_{h},{k}_{FRW})\simeq 
4\pi {n}_{0}{a}_{0}^{3}\left( \frac{{{r}_{h}}^{3}}
{3}-\frac{1}{10}{k}_{FRW}{\delta l}^{5} +O\right).
\end{equation}
The condition:
\begin{equation}
{N}_{FRW}({r}_{h},{k}_{FRW})=N({l}_{h},k(t))
\end{equation}
provides:
\begin{equation}
{N}_{0}=4\pi {n}_{0}{a}_{0}^{3}\left( \frac{{{r}_{h}}^{3}}
{3}-\frac{1}{10}{k}_{FRW}{{r}_{h}}^{5} +O\right).
\end{equation}
It should be emphasized that eq.(46) and eq.(47) allow a comparison between 
the basic FRW parameters (${a}_{0},{k}_{FRW}$) and the ones involved in the 
present model, as they should represent the large scale structure in an 
equivalent way at the homogeneity scale.\\
Although the \textit{fractal constraints} provides a link among metric
functions, it is more appealing to infer an analogous expression depending
on the luminosity distance, being it the proper observational distance
measure.\\
In this case, the same route can be followed, but an important modification
at the very beginning must be set. As discussed by other authors (Celerier
\& Thieberger 2001), if the observer area $S({d}_{a})$ expressed in terms of 
the angular distance ${d}_{a}$ and the area $S({d}_{l})$ resulting from the 
luminosity distance ${d}_{l}$ are matched, then, using eq.(36), it yields:
\begin{equation} 
S({d}_{l})=S({d}_{a}){(1+Z)}^{4}.
\end{equation}
Again, considering eq.(36) and the solid angles $\Omega ({d}_{a})$,
$\Omega ({d}_{l})$ respectively corresponding to $S({d}_{a})$, $S({d}_{l})$, 
by differentiating the above equation, an 
\textit{aberration} effect on the solid angle is found:
\begin{equation}
\Omega ({d}_{l})=\Omega ({d}_{a})\left[ 1- 4\int_{\Delta t}
{{\frac{\theta /3}{(1+Z)}}dt}\right] \equiv \Omega ({d}_{a})~
A\left( \frac{\theta}{3},Z\right) 
\end{equation}
and the \textit{aberration function} $A\left( \frac{\theta}{3},Z\right) $
introduced, as can be seen, describes the effect of the expansion on the 
subtended angle, i.e. the integration of expansion in
proper time provides the aberration of the solid angle.\\
Using such a condition, the number counts in eq.(33) can be expressed in
terms of the luminosity distance. In general, the latter is not a well
defined quantity when referring to cross sectional area perpendicular to
light rays ($\delta {l}_{0}^{2}$ in eq.(31). The actual value of 
${N}({d}_{l})$ can be obtained by inserting eq.(36) and eq.(49) in eq.(31), 
then:
\begin{equation}
{N}({d}_{l})=\frac{3{d}_{l}^{2}}{2{M}_{G}{A\left( \frac{\theta}{3},Z\right) }} 
\int_{\Delta t}{\frac{{C}^{2}(t)}{{(1+Z)}^{4}}dt}.
\end{equation}
By defining:
\begin{equation}
{I}_{\gamma }({d}_{l},Z)\equiv \int_{\Delta t}
{\frac{{C}^{2}(t)}{{(1+Z)}^{4}}dt}
\end{equation}
we get the new \textit{fractal constraint}:
\begin{equation}
{I}_{\gamma }({d}_{l},Z)=\frac{2}{3}\frac{{N}_{0}{M}_{G}}
{{l}_{h}^{2}}A\left( \frac{\theta}{3},Z\right) 
{\left( \frac{{d}_{l}}{{l}_{h}}\right) }^{1-\gamma }
\end{equation}
which indeed is now dependent on luminosity distance, redshift and
expansion, namely the observational quantities.

\section{Discussion and conclusions}

The motivation for this work has been to describe a cosmological
inhomogeneous model, which could represent a fractal 
distribution. Since spherical symmetry does not completely describe
a fractal distribution, we have considered Stephani models, i.e. inhomogeneous 
models with no intrinsic symmetries.\\
Our results are the following.\\
First, the redshift expression depends on the position; of course, this
character is present in any related cosmological parameter, such as the
expansion and the $\Omega$ value.\\
In the FRW, expansion is given by the Hubble parameter, which is the only 
term governing the Hubble law. On the converse, in the present model, 
the Hubble law has an additional dipole term due to acceleration,
which vanishes in the standard homogeneous cosmology (the quadrupole term 
due to shear vanishes as in FRW). Usually,
the observed dipole term is ascribed to peculiar velocities; in
the present model would add a pure cosmological contribution to it.\\
The existence of a kinematic acceleration in inhomogeneous models has
been invoked by other authors (Pascual-Sanchez 2000) as the origin of
accelerated expansion of the Universe, as found by recent SNae Ia data
(Perlmutter et al. 1999; Riess et al. 1999). This could be considered as
a viable alternative to the proposed interpretation within FRW scheme, namely 
a positive cosmological constant or vacuum energy or quintessence.\\
Therefore, the striking implication of the coupling of space and time 
dependences in redshift is that different 
regions of the Universe could undergo different expansion histories.\\
Actually, it should be stressed that the present model would lead to a
deeply new interpretation of observations and many cosmological and 
astrophysical contexts should be reconsidered.\\
A second result is that we have explicitly introduced a fractal 
characterization of the model.\\
The density depends on the redshift, which depends on the position. In
addition, it fulfills the scaling law imposed by the fractal distribution.
Then, the density changes according to the fractal spatial scaling law and the
redshift. This approach allows to map the density evolution on 
$t=const$ hypersurfaces, applying the fractal constraint.\\
The homogeneity scale ${l}_{h}$ plays an important role in the model.We
recall that it is the scale at which the Stephani model should match
the FRW one, i.e. the scale at which homogeneity and isotropy of the Universe 
are recovered. The greatest information supply should come by estimating
the parameters at that threshold length and imposing the FRW/fractal 
matching condition.\\
Summarizing the whole set of constraints, the present model can explicitly
express $D$, $F$ and $C$ from functions in the metric.\\
The key information still missing concerns probably the most extreme
assumption of the model $k(t)$, i.e. curvature time evolution, that is no
more kept fixed and bound to -1, 0, +1 values as in FRW Universe. Once $k(t)$
is specified, $R$ and $V$ can be fully determined
and a self-consistent cosmology can be built.\\
So far, we have listed the main differences between the FRW model and the
one proposed here. However, an important point against inhomogeneous
models is their incompatibility with high degree of isotropy of
the Cosmic Microwave Background (CMB). Indeed, it was demonstrated that an 
isotropic thermal radiation in an expanding Universe implies a FRW cosmology 
(\textit{EGS theorem}, 
Ehlers, Green \& Sachs 1968). The point is that the \textit{EGS theorem} was 
demonstrated for geodetic and non-accelerating observers, which is not the 
case in Stephani models. If acceleration is not vanishing, an 
inhomogeneous cosmology allows a thermal radiation with high degree of 
isotropy as observed (Ferrando, Morales \& Portilla 1992) and specific 
Stephani solutions can be built without any violation to the CMB restriction 
(Clarkson \& Barret 1999).\\
Nevertheless, it is to stress that limitations to the degree of
inhomogeneity of the model must be compatible with, for instance, CMB isotropy
and Hubble law local trend; for such a purpose, the derived
\textit{redshift} and \textit{fractal constraints} must be tuned.\\
Actually, a more subtle and rather philosophical argument comes into play
when facing with any inhomogeneous cosmology, that is the reliability and
interpretation of the Cosmological Principle.\\
Moreover, the present model poses serious questions about the Copernican
Principle, as well, which states that all observers are equivalent in the 
Universe. This is the basis, for instance, of the \textit{EGS theorem} and is 
far from being straightforwardly fit by a fractal structure.\\
A last comment concerns the observations, most of which can be reproduced by 
FRW cosmology. As a conservative statement, it could be noted that
each Stephani solution includes FRW as a limiting case, so that observations
cannot in general rule them out. The motivation for keeping FRW as a
description of the Universe is actually that they are the easiest
theoretical model. This led to prefer any modification to come from outside the
cosmological framework (e.g. inflation, dark matter) and not in the
cosmology itself. Therefore, the theoretical choice has been overwhelming any
observational basis, whose actual interpretation is not \textit{a priori}
excluded and could be reanalyzed within the class of inhomogeneous models.

\ack

We are grateful to Luciano Pietronero, Ruth Durrer and the PIL group for 
useful discussions. M.M. is grateful to Franco Ricci and Roberto Peron
for their continuous availability to share their expertise in GR.

\section*{References}

\smallskip
\begin{harvard}
\item[] Celerier M.-N. and Thieberger R 2001 {\it Astron.\ Astrophys.}
{\bf 367} 449
\end{harvard}
\smallskip
\smallskip
\begin{harvard}
\item[] Clarkson C A and Barret R K 1999 {\it Class.\ Quantum\ Grav.}
{\bf 16} 3781
\end{harvard}
\smallskip
\smallskip
\begin{harvard}
\item[] Coles P and Lucchin F 1995 {\it Cosmology} (New York: John Wiley
\& Sons)
\end{harvard}
\smallskip
\begin{harvard}
\item[] Ehlers J 1993 {\it Gen.\ Rel.\ Grav.} {\bf 25} 1225,
English translation from {\it Akad.\ Wiss.\ Lit.\ Moinz.\ Abhandl Math.\
Nat.\ Kl.} {\bf 11} 793 (1961)
{\bf 9} 1344
\end{harvard}
\smallskip
\smallskip
\begin{harvard}
\item[] Ehlers J, Green P and Sachs R K 1968 {\it J.\ Math.\ Phys.}
{\bf 9} 1344
\end{harvard}
\smallskip
\smallskip
\begin{harvard}
\item[] Ellis G F R, Nel S D, Maartens R, Stoeger W R and Whitman A P
1985  {\it Phys.\ Rep.} {\bf 124} 315
\end{harvard}
\smallskip
\smallskip
\begin{harvard}
\item[] Ferrando JJ, Morales J A and Portilla M 1992 {\it Phys.\ Rev.\ D} 
{\bf 46} 578
\end{harvard}
\smallskip
\smallskip
\begin{harvard}
\item[] Humphreys N P, Matravers D R and Maartens R 1998 {\it Class.\ 
Quantum\ Grav.} {\bf 15} 3781
\end{harvard}
\smallskip
\smallskip
\begin{harvard}
\item[] Joyce M, Anderson P W, Montuori M, Pietronero L, Sylos Labini F
2000 {\it Europhys.\ Lett.} {\bf 49} 416
\end{harvard}
\smallskip
\smallskip
\begin{harvard}
\item[] Krasinski A 1983 {\it Gen.\ Rel.\ and Grav.} {\bf 15} 673
\end{harvard}
\smallskip
\smallskip
\begin{harvard}
\item[] Krasinski A, Quevedo H and Sussman R A 1995 {\it J.\ Math.\ and Phys.}
{\bf 38} 20602
\end{harvard}
\smallskip
\smallskip
\begin{harvard}
\item[] Krasinski A 1997 {\it Inhomogeneous cosmological models} 
(Cambridge: Cambridge University Press)
\end{harvard}
\smallskip
\smallskip
\begin{harvard}
\item[] Mittal A K and Lohiya D 2001 {preprint} astro-ph/0104370 
\end{harvard}
\smallskip
\smallskip
\begin{harvard}
\item[] Pascual-Sanchez J-F 2000 {\it Class.\ Quantum\ Grav.} {\bf 17} 4913
\end{harvard}
\smallskip
\smallskip
\begin{harvard}
\item[] Perlmutter S et al. 1999 {\it Astrophys.\ J.} {\bf 517} 565
\end{harvard}
\smallskip
\smallskip
\begin{harvard}
\item[] Quevedo H and Sussman R A 1995 {\it J.\ Math.\ Phys.}
{\bf 36} 1365
\end{harvard}
\smallskip
\smallskip
\begin{harvard}
\item[] Ribeiro M B 1992 {\it Astrophys.\ J.} {\bf 388} 1
\end{harvard}
\smallskip
\smallskip
\begin{harvard}
\item[] Riess A G et al. 1999 {\it Astron.\ J.} {\bf 119} 1009
\end{harvard}
\smallskip
\smallskip
\begin{harvard}
\item[] Sylos Labini F, Montuori M and Pietronero L 1998 {\it Phys.\ Rep.} 
{\bf 293} 66
\end{harvard}
\smallskip
\smallskip
\begin{harvard}
\item[] Stephani H 1967 {\it Comm.\ Math.\ Phys.} {\bf 4} 137
\end{harvard}
\smallskip
\smallskip
\begin{harvard}
\item[] Wu K.K.S, Lahav O. and Rees M. 1999 {\it Nature} {\bf 397} 225
\end{harvard}
\smallskip

\end{document}